\documentclass[psfig]{mn2e} 
\usepackage{psfig}

\def\sn{\hbox{S/N}}

\def\vsin{\hbox{$v \sin i$}}  
  
\def\kms{\hbox{km\,s$^{-1}$}}  
\def\ms{\hbox{m\,s$^{-1}$}}

\def\degr{\hbox{$^\circ$}}  
\def\rpd{\hbox{rad\,d$^{-1}$}}   
\def\mrpd{\hbox{mrad\,d$^{-1}$}}   
\def\omeq{\hbox{$\Omega_{\rm eq}$}}   
\def\dom{\hbox{$d\Omega$}}   
\def\kis{\hbox{$\chi^2$}}   
\def\kisr{\hbox{$\chi^2_{\rm r}$}}   
\def\drot{\hbox{differential rotation}}   
\def\hr{\hbox{HR~1099}}

\begin{document}  

\title[Magnetic topology \& Differential rotation of HR~1099]  
{Magnetic topology and surface differential rotation on the K1 subgiant of the RS~CVn system HR~1099}
  
\makeatletter  
  
\def\newauthor{%
  \end{author@tabular}\par  
  \begin{author@tabular}[t]{@{}l@{}}}  
\makeatother  
   
\author[P.~Petit et al.]  
{\vspace{1.5mm}  
P.~Petit$^1,^2$, J.-F.~Donati$^1$, G.A.~Wade$^3$, J.D.~Landstreet$^4$, S.~Bagnulo$^5$, T.~L\"uftinger$^6$,\\ 
{\hspace{-1mm}\vspace{1.5mm}\LARGE\rm T.A.A.~Sigut$^4$, S.L.S.~Shorlin$^4$, S.~Strasser$^7$, M.~Auri\`ere$^1$, J.M.~Oliveira$^8$}\\
$^1$Laboratoire d'Astrophysique, Observatoire Midi-Pyr\'en\'ees, 14 Av.\ E.~Belin, F--31400 Toulouse, France\\ ({\tt petit@ast.obs-mip.fr, donati@ast.obs-mip.fr, auriere@ast.obs-mip.fr})\\  
$^2$Centro de Astrofisica da Universidade do Porto, rua das Estrelas, 4150-762 Porto, Portugal\\
$^3$Royal Military College of Canada, Department of Physics, P.O. Box 17000, Station "Forces", Kingston, Ontario, Canada, K7K 4B4\\ ({\tt Gregg.Wade@rmc.ca })\\
$^4$Department of Physics and Astronomy, The University of Western Ontario, London, Ontario, Canada, N6G 3K7\\ ({\tt jlandstr@astro.uwo.ca, asigut@astro.uwo.ca, sshorlin@astro.uwo.ca})\\
$^5$European Southern Observatory, Alonso de Cordova 3107, Vitacura, Santiago, 
Chile ({\tt sbagnulo@eso.org})\\
$^6$Institut f\"ur Astronomie, Tuerkenschanzstrasse 17, A-1180 Wien, Austria ({\tt theresa@tycho.astro.univie.ac.at})\\
$^7$Department of Physics and Astronomy, University of Calgary, Calgary, AB T2N 1N4, Canada ({\tt strasser@ras.ucalgary.ca})\\
$^8$Department of Physics, Keele University, Staffordshire ST5 5BG, UK ({\tt joana@astro.keele.ac.uk})\\}  
\date{2002, MNRAS}  
\maketitle  
   
\begin{abstract}   
We present here spectropolarimetric observations of the RS CVn system \hr\ (V711~Tau) secured from 1998 February to 2002 January with the spectropolarimeter MuSiCoS at the T\'elescope Bernard Lyot (Observatoire du Pic du Midi, France). We apply Zeeman-Doppler Imaging and reconstruct surface brightness and magnetic topologies of the K1 primary subgiant of the system, at five different epochs. We confirm the presence of large, axisymmetric regions where the magnetic field is mainly azimuthal, providing further support to the hypothesis that dynamo processes may be distributed throughout the whole convective zone in this star.

We study the short-term evolution of surface structures from a comparison of our images with observations secured at close-by epochs by Donati et al. (2003a) at the Anglo-Australian Telescope. We conclude that the small-scale brightness and magnetic patterns undergo major changes within a timescale of 4 to 6 weeks, while the largest structures remain stable over several years.

We report the detection of a weak surface \drot\ (both from brightness and magnetic tracers) indicating that the equator rotates faster than the pole with a difference in rotation rate between the pole and the equator about 4 times smaller than that of the Sun. This result suggests that tidal forces also impact the global dynamic equilibrium of convective zones in cool active stars.

\end{abstract}  
  

\section{Introduction}   
\label{sect:introduction}  

In classical solar dynamo theories, differential rotation plays a key role in the generation of the solar magnetic field, through its ability to transform a poloidal field into a stronger toroidal component. This so-called ``$\Omega$ effect'' is thought to take place in a thin layer at the interface between the radiative core and the convective envelope of the Sun (the tachocline). A poloidal component of the field is then regenerated by cyclonic convection (with opposite polarity), therefore producing what is known as the solar magnetic activity cycle. 

Dynamo processes generating magnetic activity on stars other than the Sun can now be directly investigated through images of their photospheric brightness and magnetic topologies, thanks to the development of Zeeman-Doppler Imaging (hereafter ZDI, Donati \& Brown 1997). The long-term monitoring of a sample of active fast rotators shows that they all possess regions of strong, nearly azimuthal magnetic field, features that are not expected to show up at photospheric level in the context of the conventional solar dynamo. Such observations suggest that the $\Omega$ effect is not confined in the tachocline of these objects, but may be instead distributed throughout their convective envelope, and at least be efficient close to the photosphere.

In addition to yielding information about the magnetic topology, indirect imaging techniques provide the first opportunity for monitoring the short-term evolution of brightness and magnetic inhomogeneities at the surface of active stars, under the influence of differential rotation. The first results, derived from observations of several young fast rotators (Donati \& Cameron 1997, Donati et al. 1999, Donati et al. 2000, Barnes et al. 2000, Cameron et al. 2002, Donati et al 2002b) show that their surface rotational shear is of the same magnitude as that of the Sun, with a trend to increase toward higher stellar masses (as confirmed by the observations of Reiners \& Schmitt 2002 on a sample of F-type fast rotators).

The K1 primary subgiant of the RS~CVn system \hr\ is one of the most active stars in the whole sky. The complex surface magnetic field of this object has already been mapped by Donati et al. (1992), Donati (1999) and Donati et al. (2003a), confirming that efficient dynamo processes are operating within its deep convective envelope. Unfortunately, the first attempts at estimating \drot\ at the surface of this object (Vogt et al. 1999, Strassmeier \& Bartus 2000) were not convincing because they were based on data of relatively low \sn\ and sparse phase sampling (Petit et al. 2002). However, a rigorous analysis of the surface \drot\ of this star should provide useful information complementary to magnetic imaging, for investigating the dynamo mechanisms driving its impressive activity. Finally, the present study is of special interest as it concerns a star with different evolutionary status than those studied to date, featuring in particular a very deep convective zone and subject to intense tidal forces.

In the present paper, we report new spectropolarimetric observations of \hr\ secured between 1998 and 2002 at Observatoire du Pic du Midi (France), from which we reconstruct brightness and magnetic images of its primary subgiant by means of ZDI. Comparisons are made with observations of the same object secured by Donati et al. 2003a (D03a from now on) at the AAT a few weeks apart from, and sometimes simultaneously with, our own observations. We take the original opportunity of comparing images obtained with different instrumental setups and observing conditions for providing further testing of the robustness of ZDI. We also study the short-term evolution of surface structures by comparison with observations of D03a obtained within a few weeks from our own data. 

We then apply the method of Petit et al. (2002) to estimate the surface rotational shear from brightness and magnetic images. We compare our results with the work of Donati et al. 2003b (hereafter D03b), who derived a first estimate of differential rotation on this star using the same method, but from data sets substantially smaller than ours. We also discuss the possibility of detecting temporal fluctuations of the shear, as reported by D03b for younger objects.

We finally summarize the results and investigate the possible connection of the measured surface shear with the observed lifetime of surface structures, as well as the possible impact of tidal forces on the differential rotation of \hr.

\section{Observations and data preparation} 
\label{sect:observations} 

\subsection{Data collection and reduction}

\begin{table*}
\caption[]{Journal of observations for epoch 1998.14. Each line corresponds to a full polarization cycle. The two figures separated by a ``/'' give the minimum and maximum value of each field, except in column 4, listing the number of unpolarized/polarized spectra. Column 5 lists the total exposure time of each Stokes I individual sub-exposure. We also list the \sn\ ratios (per 4 \kms\ velocity bins) of the unpolarized and polarized spectra (in columns 6 and 8 respectively) and in the associated mean LSD profiles (columns 7 and 9). The multiplex gain between the raw polarized spectra and the mean Stokes V profiles is reported in the last column.}
\begin{tabular}{cccccccccc}
\hline
Date  & JD           & UT & nexp & t$_{\rm exp}$ & \sn & \sn          & \sn  & \sn          & multiplex gain \\
      & (+2,450,000) & (hh$:$mm$:$ss)   &      & (sec.)        &  I  & I$_{\rm LSD}$ & V    & V$_{\rm LSD}$ & V \\
\hline
1998 Feb 4 & 849.317/849.332 & 19:35:43/19:58:22 & 4/1 & 360 & 160/170 & 864/873 & 330 & 12020 & 36\\
1998 Feb 5 & 850.311/850.335 & 19:27:27/20:01:50 & 5/1 & 360 & 130/140 & 958/964 & 270 & 9081 & 33\\
1998 Feb 6 & 851.307/851.323 & 19:22:30/19:45:22 & 4/1 & 360 & 140/150 & 957/960 & 290 & 9719 & 33\\
1998 Feb 7 & 852.296/852.328 & 19:06:00/19:53:00 & 5/1 & 600 & 130/150 & 939/956 & 310 & 10133 & 32\\
1998 Feb 8 & 853.324/853.340 & 19:46:48/20:09:00 & 4/1 & 360 & 110/140 & 827/844 & 250 & 8023 & 32\\
1998 Feb 9 & 854.345/854.363 & 20:16:49/20:43:26 & 4/1 & 360 & 160/170 & 958/965 & 330 & 10516 & 31\\
1998 Feb 10 & 855.377/855.394 & 21:03:00/21:26:45 & 4/1 & 360 & 140/150 & 984/993 & 300 & 9523 & 31\\
1998 Feb 11 & 856.341/856.357 & 20:11:03/20:33:36 & 4/1 & 360 & 160/170 & 845/848 & 330 & 11178 & 33\\
1998 Feb 12 & 857.301/857.317 & 19:13:35/19:35:43 & 4/1 & 360 & 160/180 & 954/958 & 350 & 11989 & 34\\
1998 Feb 13 & 858.316/858.337 & 19:35:11/20:05:00 & 5/1 & 360 & 150/170 & 981/985 & 360 & 12351 & 34\\
1998 Feb 14 & 859.326/859.339 & 19:49:11/20:07:34 & 4/1 & 300 & 110/120 & 884/893 & 230 & 7101 & 30\\
1998 Feb 15 & 860.337/860.351 & 20:05:35/20:26:02 & 4/1 & 300 & 87/90 & 908/915 & 170 & 4781 & 28\\
1998 Feb 16 & 861.339/861.353 & 20:08:33/20:28:09 & 4/1 & 300 & 110/120 & 978/982 & 240 & 8275 & 34\\
1998 Feb 17 & 862.310/862.323 & 19:26:04/19:45:27 & 4/1 & 300 & 100/110 & 910/927 & 210 & 6505 & 30\\
1998 Feb 18 & 863.291/863.307 & 18:59:25/19:21:42 & 4/1 & 360 & 120/140 & 900/910 & 250 & 7986 & 31\\
1998 Feb 20 & 865.319/865.342 & 19:39:32/20:13:05 & 4/1 & 600 & 92/120 & 937/949 & 210 & 6600 & 31\\
1998 Feb 26 & 871.353/871.369 & 20:28:14/20:50:56 & 4/1 & 360 & 100/110 & 937/954 & 200 & 5657 & 28\\
1998 Feb 27 & 872.328/872.343 & 19:51:50/20:14:20 & 4/1 & 360 & 140/140 & 898/907 & 280 & 8536 & 30\\
1998 Feb 28 & 873.359/873.375 & 20:36:40/20:59:23 & 4/1 & 360 & 120/150 & 842/845 & 270 & 7715 & 28\\
1998 Mar 2 & 875.348/875.364 & 20:20:50/20:44:10 & 4/1 & 360 & 100/120 & 940/964 & 220 & 6189 & 28\\
1998 Mar 5 & 878.377/878.394 & 21:02:58/21:27:29 & 4/1 & 360 & 95/100 & 528/931 & 190 & 3514 & 18\\
\hline
\end{tabular}
\label{tab:journalfeb98}
\end{table*}

\begin{table*}
\caption[]{Same as Table \ref{tab:journalfeb98} for epochs 1998.93 and 1999.06.}
\begin{tabular}{cccccccccc}
\hline
Date  & JD           & UT & nexp & t$_{\rm exp}$ & \sn & \sn          & \sn  & \sn          & multiplex gain \\
      & (+2,450,000) & (hh$:$mm$:$ss)   &      & (sec.)        &  I  & I$_{\rm LSD}$ & V    & V$_{\rm LSD}$ & V \\
\hline
1998 Dec 5 & 1153.3298/1153.3528 & 19:54:56/20:27:58 & 4/1 & 360 & 96/102 & 930/977 & 195 & 5912 & 30\\
1998 Dec 5 & 1153.4280/1153.4445 & 22:16:22/22:40:04 & 4/1 & 360 & 95/122 & 978/979 & 230 & 7458 & 32\\
1998 Dec 6 & 1153.5307/1153.5492 & 00:44:15/01:10:50 & 4/1 & 420 & 127/154 & 973/987 & 275 & 8792 & 32\\
1998 Dec 6 & 1153.6077/1153.6237 & 02:35:03/02:58:11 & 4/1 & 360 & 68/88 & 923/964 & 159 & 4396 & 28\\
1998 Dec 7 & 1155.4395/1155.4554 & 22:32:49/22:55:44 & 4/1 & 360 & 92/125 & 854/932 & 224 & 7036 & 31\\
1998 Dec 8 & 1155.5910/1155.6070 & 02:11:04/02:34:09 & 4/1 & 360 & 97/112 & 836/864 & 206 & 5543 & 27\\
1998 Dec 8 & 1156.3413/1156.3572 & 20:11:27/20:34:22 & 4/1 & 360 & 47/60 & 822/862 & 106 & 2981 & 28\\
\hline
1999 Jan 13 & 1192.2812/1192.2958 & 18:44:54/19:06:00 & 4/1 & 360 & 103/127 & 878/909 & 226 & 6885 & 30\\
1999 Jan 14 & 1193.3795/1193.3943 & 21:06:26/21:27:46 & 4/1 & 360 & 105/127 & 942/954 & 232 & 7152 & 31\\
1999 Jan 19 & 1198.3154/1198.3303 & 19:34:10/19:55:34 & 4/1 & 360 & 131/148 & 827/840 & 280 & 9278 & 33\\
1999 Jan 22 & 1201.3794/1201.3942 & 21:06:21/21:27:38 & 4/1 & 360 & 142/148 & 933/944 & 284 & 8837 & 31\\
1999 Jan 23 & 1202.3167/1202.3315 & 19:36:07/19:57:22 & 4/1 & 360 & 165/178 & 893/921 & 344 & 11339 & 34\\
1999 Jan 24 & 1203.3362/1203.3557 & 20:04:00/20:32:11 & 4/1 & 360 & 116/125 & 898/907 & 238 & 7347 & 31\\
1999 Jan 25 & 1204.3249/1204.3399 & 19:47:48/20:09:24 & 4/1 & 360 & 99/110 & 906/942  & 205 & 6306 & 31\\
1999 Jan 30 & 1209.3488/1209.3634 & 20:22:20/20:43:20 & 4/1 & 360 & 45/48 & 540/602  & 93 & 2504 & 27\\
1999 Jan 31 & 1210.2966/1210.3114 & 19:07:10/19:28:22 & 4/1 & 360 & 93/108 & 833/897 & 198 & 5995 & 30\\
1999 Jan 31 & 1210.3627/1210.3814 & 20:42:14/21:09:10 & 4/1 & 480 & 116/128 & 890/915 & 242 & 7486 & 31\\
\hline
\end{tabular}
\label{tab:journaldec98jan99}
\end{table*}

\begin{table*}
\caption[]{Same as Table \ref{tab:journalfeb98} for epoch 2000.14.}
\begin{tabular}{cccccccccc}
\hline
Date  & JD           & UT & nexp & t$_{\rm exp}$ & \sn & \sn          & \sn  & \sn          & multiplex gain \\
      & (+2,450,000) & (hh$:$mm$:$ss)   &      & (sec.)        &  I  & I$_{\rm LSD}$ & V    & V$_{\rm LSD}$ & V \\
\hline
2000 Feb 3 & 1578.3474/1578.3738 & 20:20:15/20:58:15 & 6/1 & 360 & 83/97 & 861/877 & 170 & 5062 & 29 \\
2000 Feb 4 & 1579.3538/1579.3688 & 20:29:24/20:51:01 & 4/1 & 360 & 66/76 & 882/915 & 130 & 4251 & 32 \\
2000 Feb 5 & 1580.3578/1580.3779 & 20:35:16/21:04:00 & 4/1 & 360 & 38/67 & 534/773 & 230 & 6618 & 28 \\
2000 Feb 9 & 1584.3295/1584.3441 & 19:54:30/20:15:30 & 4/1 & 360 & 110/130 & 917/932 & 100 & 2326 & 23 \\
2000 Feb 11 & 1586.3744/1586.3890 & 20:59:10/21:20:10 & 4/1 & 360 & 98/100 & 920/956 & 200 & 5089 & 25 \\
2000 Feb 12 & 1587.3340/1587.3486 & 20:00:56/20:21:56 & 4/1 & 360 & 130/140 & 971/991 & 270 & 8069 & 29 \\
2000 Feb 15 & 1590.3453/1590.3599 & 20:17:10/20:38:13 & 4/1 & 360 & 100/110 & 982/994 & 210 & 5812 & 27 \\
2000 Feb 22 & 1597.3141/1597.3290 & 19:32:21/19:53:43 & 4/1 & 360 & 140/150 & 975/980 & 280 & 8352 & 29 \\
2000 Feb 24 & 1599.3157/1599.3303 & 19:34:37/19:55:39 & 4/1 & 360 & 110/120 & 994/1007 & 220 & 6015 & 27 \\
2000 Feb 25 & 1600.3060/1600.3206 & 19:20:40/19:41:37 & 4/1 & 360 & 130/150 & 975/987 & 270 & 8213 & 30 \\
2000 Feb 26 & 1601.3236/1601.3383 & 19:45:56/20:07:00 & 4/1 & 360 & 130/100 & 893/904 & 220 & 6249 & 28 \\
2000 Feb 27 & 1602.3092/1602.3241 & 19:25:17/19:46:40 & 4/1 & 360 & 100/120 & 969/983 & 220 & 6434 & 29 \\
2000 Mar 2 & 1606.3150/1606.3301 & 19:33:32/19:55:24 & 4/1 & 360 & 120/140 & 992/1007 & 270 & 7530 & 27 \\
2000 Mar 3 & 1607.3187/1607.3345 & 19:38:57/20:01:39 & 3/0 & 600 & 49/60 & 618/761 & -- & -- & -- \\
2000 Mar 4 & 1608.2950/1608.3100 & 19:04:50/19:26:26 & 4/1 & 360 & 140/150 & 872/879 & 290 & 8249 & 28 \\
2000 Mar 5 & 1609.2978/1609.3125 & 19:08:53/19:30:04 & 4/1 & 360 & 92/100 & 769/865 & 180 & 5059 & 28 \\
2000 Mar 8 & 1612.2965/1612.3113 & 19:06:57/19:28:17 & 4/1 & 360 & 110/130 & 798/948 & 230 & 7319 & 31 \\
\hline
\end{tabular}
\label{tab:journalfeb00}
\end{table*}

\begin{table*}
\caption[]{Same as Table \ref{tab:journalfeb98} for epoch 2001.96.}
\begin{tabular}{cccccccccc}
\hline
Date  & JD           & UT & nexp & t$_{\rm exp}$ & \sn & \sn          & \sn  & \sn          & multiplex gain \\
      & (+2,450,000) & (hh$:$mm$:$ss)   &      & (sec.)        &  I  & I$_{\rm LSD}$ & V    & V$_{\rm LSD}$ & V \\
\hline
2001 Dec 1 & 2245.4601/2245.4756 & 23:02:32/23:24:53 & 4/1 & 400 & 119/142 & 888/890 & 265 & 8428 & 31 \\
2001 Dec 2 & 2245.5809/2245.5964 & 01:56:32/02:18:53 & 4/1 & 400 & 117/124 & 908/917 & 243 & 7125 & 29 \\
2001 Dec 2 & 2246.3264/2246.3419 & 19:49:59/20:12:21 & 4/1 & 400 & 106/114 & 942/956 & 217 & 6589 & 30 \\
2001 Dec 2 & 2246.4399/2246.4554 & 22:33:29/22:55:50 & 4/1 & 400 & 106/156 & 948/960 & 260 & 8160 & 31 \\
2001 Dec 3 & 2246.5517/2246.5673 & 01:14:30/01:36:52 & 4/1 & 400 & 142/157 & 913/924 & 303 & 9354 & 30 \\
2001 Dec 5 & 2249.3259/2249.3415 & 19:49:22/20:11:43 & 4/1 & 400 & 63/86 & 869/906 & 148 & 4301 & 28 \\
2001 Dec 5 & 2249.4378/2249.4533 & 22:30:24/22:52:46 & 4/1 & 400 & 69/95 & 692/792 & 158 & 4622 & 29 \\
2001 Dec 6 & 2249.5052/2249.5207 & 00:07:25/00:29:46 & 4/1 & 400 & 52/64 & 519/612 & 112 & 3092 & 27 \\
2001 Dec 6 & 2250.3536/2250.3692 & 20:29:15/20:51:36 & 4/1 & 400 & 110/115 & 909/924 & 251 & 7893 & 31 \\
2001 Dec 6 & 2250.4447/2250.4603 & 22:40:26/23:02:48 & 4/1 & 400 & 121/124 & 919/944 & 215 & 5498 & 25 \\
2001 Dec 7 & 2250.5738/2250.5894 & 01:46:18/02:08:40 & 4/1 & 400 & 117/123 & 915/930 & 222 & 6776 & 30 \\
2001 Dec 7 & 2251.3256/2251.3411 & 19:48:52/20:11:13 & 4/1 & 400 & 124/136 & 936/951 & 244 & 7658 & 31 \\
2001 Dec 7 & 2251.4368/2251.4524 & 22:29:02/22:51:24 & 4/1 & 400 & 124/130 & 941/954 & 236 & 6713 & 28 \\
2001 Dec 8 & 2251.5824/2251.5979 & 01:58:39/02:21:01 & 4/1 & 400 & 105/108 & 918/947 & 279 & 8972 & 32 \\
2001 Dec 8 & 2252.3268/2252.3423 & 19:50:34/20:12:55 & 4/1 & 400 & 123/127 & 860/869 & 281 & 8961 & 31 \\
2001 Dec 8 & 2252.4595/2252.4750 & 23:01:37/23:23:58 & 4/1 & 400 & 110/124 & 828/846 & 282 & 7662 & 27 \\
2001 Dec 9 & 2252.5721/2252.5876 & 01:43:49/02:06:11 & 4/1 & 400 & 115/117 & 877/888 & 260 & 8243 & 31 \\
2001 Dec 9 & 2253.4321/2253.4476 & 22:22:10/22:44:31 & 4/1 & 400 & 124/127 & 949/958 & 252 & 7928 & 31 \\
2001 Dec 10 & 2253.5768/2253.5923 & 01:50:38/02:12:59 & 4/1 & 400 & 111/118 & 928/932 & 212 & 5772 & 27 \\
2001 Dec 10 & 2254.3562/2254.3717 & 20:32:55/20:55:16 & 4/1 & 400 & 130/137 & 975/984 & 248 & 7833 & 31 \\
2001 Dec 10 & 2254.4663/2254.4818 & 23:11:28/23:33:49 & 4/1 & 400 & 128/131 & 972/980 & 231 & 7159 & 30 \\
2001 Dec 11 & 2254.5791/2254.5946 & 01:53:52/02:16:13 & 4/1 & 400 & 124/131 & 964/979 & 232 & 6487 & 27 \\
2001 Dec 11 & 2255.3328/2255.3483 & 19:59:12/20:21:34 & 4/1 & 400 & 119/131 & 846/857 & 249 & 7869 & 31 \\
2001 Dec 11 & 2255.4660/2255.4815 & 23:10:59/23:33:21 & 4/1 & 400 & 109/131 & 901/919 & 231 & 6324 & 27 \\
2001 Dec 12 & 2255.5897/2255.6052 & 02:09:00/02:31:28 & 4/1 & 400 & 78/102 & 875/921 & 268 & 8339 & 31 \\
2001 Dec 12 & 2256.3289/2256.3445 & 19:53:40/20:16:02 & 4/1 & 400 & 134/144 & 945/949 & 259 & 8183 & 31 \\
2001 Dec 12 & 2256.4443/2256.4599 & 22:39:51/23:02:12 & 4/1 & 400 & 148/155 & 919/934 & 251 & 7000 & 27 \\
2001 Dec 13 & 2256.5549/2256.5704 & 01:19:00/01:41:21 & 4/1 & 400 & 132/136 & 881/891 & 250 & 7747 & 30 \\
2001 Dec 13 & 2257.3309/2257.3465 & 19:56:34/20:18:55 & 4/1 & 400 & 195/200 & 825/878 & 236 & 7361 & 31 \\
2001 Dec 13 & 2257.4459/2257.4614 & 22:42:03/23:04:25 & 4/1 & 400 & 187/190 & 733/834 & 180 & 4916 & 27 \\
2001 Dec 14 & 2257.5264/2257.5419 & 00:37:59/01:00:20 & 4/1 & 400 & 174/176 & 696/751 & 276 & 8494 & 30 \\
2001 Dec 16 & 2259.5505/2259.5661 & 01:12:46/01:35:08 & 4/1 & 400 & 61/78 & 816/842 & 303 & 9797 & 32 \\
2001 Dec 16 & 2260.3293/2260.3448 & 19:54:10/20:16:31 & 4/1 & 400 & 152/159 & 972/978 & 270 & 7972 & 29 \\
2001 Dec 16 & 2260.4268/2260.4423 & 22:14:36/22:36:58 & 4/1 & 400 & 126/159 & 947/967 & 291 & 9142 & 31 \\
2001 Dec 17 & 2260.5529/2260.5684 & 01:16:11/01:38:33 & 4/1 & 400 & 123/127 & 896/937 & 261 & 8311 & 31 \\
2001 Dec 17 & 2261.2953/2261.3108 & 19:05:13/19:27:36 & 4/1 & 400 & 134/146 & 939/954 & 185 & 5425 & 29 \\
2001 Dec 17 & 2261.4359/2261.4515 & 22:27:46/22:50:09 & 4/1 & 400 & 144/146 & 932/948 & 140 & 3731 & 26 \\
2001 Dec 18 & 2261.5424/2261.5580 & 01:01:00/01:23:27 & 4/1 & 400 & 131/141 & 914/938 & 310 & 10226 & 32 \\
2001 Dec 18 & 2262.2848/2262.3004 & 18:50:08/19:12:31 & 4/1 & 400 & 134/154 & 856/858 & 294 & 9494 & 32 \\
2001 Dec 18 & 2262.4283/2262.4438 & 22:16:42/22:39:05 & 4/1 & 400 & 120/157 & 845/852 & 249 & 7105 & 28 \\
2001 Dec 19 & 2262.5537/2262.5692 & 01:17:19/01:39:41 & 4/1 & 400 & 115/128 & 823/862 & 280 & 8646 & 30 \\
2001 Dec 20 & 2264.2891/2264.3046 & 18:56:17/19:18:40 & 4/1 & 400 & 94/105 & 917/930 & 289 & 9152 & 31 \\
2001 Dec 21 & 2265.2786/2265.2941 & 18:41:07/19:03:30 & 4/1 & 400 & 88/135 & 858/877 & 274 & 7798 & 28 \\
2001 Dec 22 & 2266.2873/2266.3028 & 18:53:43/19:16:06 & 4/1 & 400 & 97/103 & 924/939 & 291 & 9180 & 31 \\
2001 Dec 27 & 2271.4947/2271.5169 & 23:52:26/00:24:22 & 4/1 & 400 & 95/138 & 955/985 & 291 & 9476 & 32 \\
2002 Jan 5 & 2280.3670/2280.3825 & 20:48:26/21:10:49 & 4/1 & 400 & 124/127 & 941/952 & 242 & 6637 & 27 \\
2002 Jan 6 & 2280.5197/2280.5353 & 00:28:25/00:50:48 & 4/1 & 400 & 101/112 & 716/919 & 200 & 5871 & 29 \\
2002 Jan 6 & 2281.2897/2281.3053 & 18:57:12/19:19:34 & 4/1 & 400 & 136/142 & 936/950 & 230 & 7065 & 30 \\
2002 Jan 6 & 2281.3785/2281.3941 & 21:05:00/21:27:29 & 4/1 & 400 & 138/144 & 873/937 & 200 & 5043 & 25 \\
2002 Jan 7 & 2281.5005/2281.5161 & 00:00:44/00:23:07 & 4/1 & 400 & 137/146 & 773/811 & 252 & 7473 & 29 \\
\hline
\end{tabular}
\label{tab:journaldec01}
\end{table*}

The polarized spectral data used in this paper were secured with T\'elescope Bernard Lyot 
(TBL) at Observatoire du Pic du Midi (France) between February 1998 and January 
2002, using the MuSiCoS echelle spectrograph (Baudrand \& Boehm 1992) fibre 
fed from a Cassegrain-mounted polarimetric module (Donati et al. 
1999). The data reduction was performed with ESpRIT, following the procedure developed by Donati et al. (1997). 

The journal of observations is reported in Tables \ref{tab:journalfeb98} to \ref{tab:journaldec01}. We divide the observations into 5 separate groups, corresponding to different observing periods (epoch 1998.14 in Table \ref{tab:journalfeb98}, epochs 1998.93 and 1999.06 in Table \ref{tab:journaldec98jan99}, epoch 2000.14 in Table \ref{tab:journalfeb00} and epoch 2001.96 in Table \ref{tab:journaldec01}). This observing effort yielded a total of 104 polarized exposures secured over 72 telescope nights. Exposures in circular polarization (Stokes V) consist of a sequence of 4 sub-exposures taken with the quarter-wave plate oriented at azimuth $\pm 45 \degr$ with respect to the optical axis of the beam-splitter. Each of these individual sub-exposures are used to produce an intensity spectrum (Stokes I), thus allowing us to take advantage of the finer phase resolution (of the order of $1.5\times 10^{-3}$ rotation cycle for an exposure of 360 seconds) between successive sub-exposures. The total number of Stokes I spectra reported in this study is 420. 

Least Square Deconvolution (LSD, Donati et al. 1997) is employed 
to extract simultaneously the information from most spectral lines available in the observed wavelength domain. The spectral 
line list used to produce the LSD profiles of HR~1099 corresponds to the same K1 photospheric 
mask as that employed in Donati (1999). A total of 
about 2,600 spectral features are taken into account for the present data sets 
as opposed to about 4500 in the study of Donati et al. 2002a. This difference results from the smaller spectral range of MuSiCoS (450 to 660 nm in polarimetric mode), which reduces the number of spectral lines available for LSD. The resulting Stokes V profiles benefit from a multiplex gain of about 30 (see Tables \ref{tab:journalfeb98} to \ref{tab:journaldec01}) as opposed to an average 
multiplex gain of 40 for AAT profiles. Most Stokes I LSD profiles \sn\ values stay between 900 and 1000, following the behavior pointed out by Donati et al. (1997), and indicating that the convolution model underlying LSD cannot be trusted above this level of accuracy. 

The wavelength calibration, automatically performed by ESpRIT with a reference 
to a Thorium-Argon spectrum, has been refined for the present study by means of a new procedure involving telluric lines, and is similar to that described in D03a. This method consists in running LSD on each stellar spectrum with a special mask including telluric lines only. The measured shift of the derived LSD telluric profiles from a null velocity (in the terrestrial frame) is attributed to instrumental instability and used to correct the wavelength scale. This method permits an exposure-by-exposure calibration, thus suppressing the instrumental drifts occurring throughout the nights. The accuracy of the telluric lines calibration (which can roughly be 
evaluated e.g. by the average wavelength shift between two successive exposures) 
is of order of 300 \ms, which is not as good a value as that reached with 
UCLES at the AAT ($< 100 \ms$). Most of the difference can be attributed to the smaller 
spectral range of TBL (especially in the red region), which reduces the number 
of telluric lines available, as well as to the site quality of Observatoire du Pic du Midi (where the telluric spectrum is usually weak, thanks to the altitude of the telescope). 
Applying this technique shows that 
the wavelength shift across the night can be either slow and progressive or abrupt and 
very fast (i.e. between two successive exposures separated by a few minutes), reaching as much as 2 \kms, i.e. more than 6 times the accuracy of telluric line calibration. We thus conclude that such wavelength calibration is significantly better than that achieved from ThAr spectra alone.

\subsection{Imaging procedure}
\label{sect:procedure}

\begin{table}
\caption[]{Orbital parameters of HR~1099 estimated from our data sets. $K_s$, $\gamma$ and $\phi_0$ are respectively the radial velocity amplitude of the secondary star, the radial velocity shift of the system and the first conjunction phase of the system. Error bars are of order of $0.3$ \kms for $K_s$ and $\gamma$, and $2\times10^{-4}$ for $\phi_0$.}
\centerline{\begin{tabular}{cccc}
Epoch  &   $K_s$   &   $\gamma$ & $\phi_0$  \\
  &   \kms  &  \kms  &         \\
\hline
1998.14 & 62.80 & --14.2  & --0.0452 \\
1998.93 & 62.32 & --13.8  & --0.0513 \\
1999.06 & 62.66 & --14.5  & --0.0541 \\
2000.14 & 62.85 & --14.2  & --0.0615 \\
2001.96 & 62.82 & --14.6  & --0.0791 \\ 
\hline
\end{tabular}}
\label{tab:orbit}
\end{table}

All the brightness and magnetic images described in this paper are obtained with the imaging code developed by Brown et al. (1991) and Donati \& Brown (1997), following the principles of maximum entropy image reconstruction outlined by Skilling \& Bryan (1984). The behavior of this imaging procedure was tested for various stellar parameters and observing conditions by Donati \& Brown (1997), from a series of numerical simulations. They demonstrated that regions in which the field orientation is azimuthal can clearly be distinguished from radial or meridional field structures (for noise levels similar to that available for the present study). Some of their conclusions should however be kept in mind in the specific case of \hr, as outlined by Donati (1999). In particular, we expect a partial crosstalk between radial and meridional field components located at low latitudes, owing to the relatively low inclination angle of the star. Moreover, in case of images computed with an incomplete phase sampling, only a partial reconstruction of the magnetic field is achieved, containing radial/meridional field regions closest to the observed longitudes and azimuthal field structures located about 0.2 rotation cycle away from the observed longitudes. 

To model the photospheric brightness inhomogeneities, we use the two component description of Cameron et al. (1992), in which every pixel of the stellar surface includes a fraction $f$ of quiet photosphere (of temperature 4750 K) and $1 - f$ of cool spot (of temperature 3500 K). The average intrinsic profile used for modeling the observed LSD spectra and compute brightness images is a synthetic Gaussian line reproducing the characteristics of a MuSiCoS LSD Stokes I profile of the K0 star $\beta$~Gem. This option was adopted according to the results of Unruh \& Cameron (1995), who demonstrated that Doppler images reconstructed from a Gaussian line were almost indistinguishable from that obtained using a standard star. The use of a synthetic line further guarantees that the template is free from noise. This Gaussian profile was scaled by a factor 0.5 and 1 for the spotted areas and the quiet photosphere respectively, as suggested by the observations of Donati \& Cameron (1997).

Before reconstructing brightness and magnetic images of the primary component of HR~1099, one must first perform a careful removal of the contribution from the G5 secondary component to the system spectra, following the method outlined in Donati et al. (1992). This correction consists in suppressing the line of the secondary star (assumed Gaussian) in every Stokes I profiles of the system at conjunction phases. We assume that the contribution of the secondary star can be neglected in Stokes V profiles, as discussed by Donati (1999). Then one has to correct for the orbital motion of the primary star. Both operations require a precise estimate of the system orbital parameters. 

The radial velocity amplitude $K_s$ of the secondary star, the radial mean velocity shift $\gamma$ of the system and the phase of conjunction $\phi_0$ of both stars are listed in Table \ref{tab:orbit}. These values are in very good agreement with estimates derived from AAT data at close-by epochs. We thus confirm the stability of $K_s$ and $\gamma$ over years, and the persistent decrease of $\phi_0$, by an average of $0.89\%$ of a rotation cycle in one year. The determination of the velocity amplitude \vsin\ and inclination $i$ of the primary star is discussed by Donati (1999). In this paper, we set these parameters to 40 \kms and 38\degr\ respectively (slightly different from, though still compatible with, Donati 1999), to provide the best fit to our data.   

\section{Reconstructed images}  
\label{sect:images}  

The 5 brightness and magnetic images obtained for the different observing epochs are described in this section. In order to allow easy comparisons (and keep consistent) with previous works presenting ZDI images of \hr\ (Donati 1999, D03a), the reconstructed magnetic field is divided into its three components in spherical coordinates, each one displayed in a gray-scale chart. In recent studies, some authors rather display 3D field lines in one single chart (see, e.g., Piskunov 2001), with the obvious advantage that the field orientation is therefore directly seen on the map. This second option may be well-adapted to the case of chemically peculiar stars, hosting a large-scale field varying smoothly over the stellar photosphere. On the surface of cool active stars however, individual magnetic regions cover only a small fraction of the photosphere and the orientation of field lines can vary a lot from one region to the other. In this specific case, 3D field lines would therefore produce a tangled pattern, very hard (if not impossible) to interpret. Displaying 3 distinct maps appears to be a good compromise in this context, but the reader should keep in mind that the exact orientation of field lines inside active regions can only be deduced from a comparison of all three sub-components of the magnetic field.

\begin{figure*}  
\centerline{\psfig{file=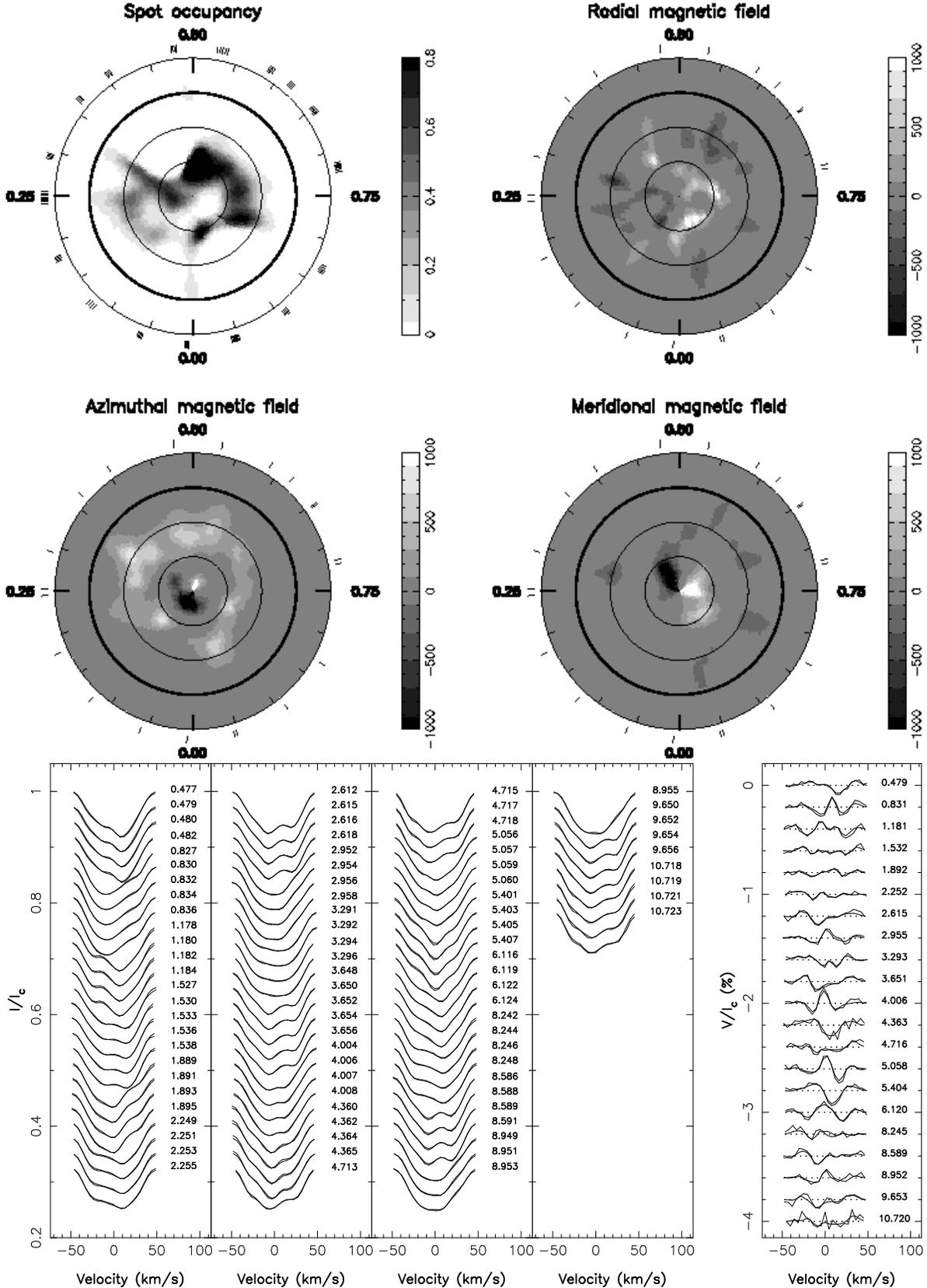,height=13cm}}
\centerline{\mbox{\psfig{file=specI_feb98.ps,height=9cm,angle=270}  
\hspace{2mm}      \psfig{file=specV_feb98.ps,height=9cm,angle=270}}}
\caption[]{Reconstructed images of the primary star of HR~1099 at epoch 1998.14, in flattened polar view. The concentric circles correspond (from outer side to center) to parallels of latitude $-30$\degr, 0\degr (equator, bold line), $+30$\degr and $+60$\degr. The upper-left panel corresponds to a brightness image, while the three other panels show the components of the magnetic field (in Gauss) in spherical coordinates, i.e. radial, azimuthal and meridional components of the field in the upper-right, lower-left and lower-right panels respectively. Mean Stokes I and V profiles (left-hand panel and right-hand panel, respectively) of the primary star of HR~1099 for the 1998.14 data set are also depicted in the bottom graph. Thin lines represent the observed profiles, while bold lines correspond to the profiles reconstructed by the imaging code.}  
\protect\label{fig:feb98}  
\end{figure*}  

\subsection{Phase sampling and timespan of observations}
\label{sampling}

Three of the data sets studied in this paper present a time span of observations of order one month (30 nights at epoch 1998.14, 35 nights for 2000.14 and 37 nights for 2001.96). The rather substantial number of Stokes I and V profiles constituting these data sets (200 Stokes I / 50 Stokes V profiles for epoch 2001.96, 65/15 for 2000.14, 87/21 for 1998.14, see Tables \ref{tab:journaldec01}, \ref{tab:journalfeb00} and \ref{tab:journalfeb98}) provides a very dense (and sometimes even redundant) phase coverage (for epochs 1998.93 and 1999.06, large phase gaps only allow a partial reconstruction of the brightness and magnetic topologies). However, the drawback of this long time interval of data collection is that the images we present are far from being, as they ideally should be, a snapshot of the surface of HR~1099. In this respect, they are susceptible to be affected by long-term changes occurring during the data collection. 

Some of the temporal changes in the photospheric distribution can be dealt with by the imaging procedure as long as they can be accurately modeled. This can be done for instance in the case of large-scale surface flows, for which the time dependence of surface motion follows simple laws. For instance, differential rotation is implemented within the imaging code, assuming a rotation law of the form~:

\begin{equation}
\Omega(l) = \omeq - \dom \sin^2 l
\label{eq:diffrot}
\end{equation}

\noindent where $\Omega(l)$ is the rotation rate at latitude $l$, \omeq\ the rotation rate of the equator and \dom\ the difference in rotation rate between the pole and the equator. Section \ref{sect:diffrot} reports our attempt at using this technique to derive \drot\ at the surface of \hr\ from the present data set. All the images described hereafter are reconstructed assuming $\omeq = 2.222$ \rpd\ and $\dom = 17$ \mrpd, following the preliminary estimate of Petit et al. (2001)\footnote{where the \drot\ parameters were by mistake given in $\rm rad~s^{-1}$, instead of \rpd.}.  

More problematic are the local (and unpredictable) modifications of surface structures, such as appearing or vanishing features. This kind of short term evolution will be discussed in Sect. \ref{sect:short} by comparing our images with those reconstructed from AAT data sets secured typically a few weeks apart from our observations. 

The first indication for the temporal variability of surface structures on \hr\ during the observations comes from the monitoring of the reduced \kis\ (hereafter \kisr) of the reconstructed profiles (see Table \ref{tab:kisr}). The \kisr\ of magnetic images ranges from 0.9 to 1.3, while brightness images are reconstructed with a \kisr\ ranging from 1.1 to 1.4 (leaving out the particular case of the images obtained by grouping data obtained at epochs 1998.93 and 1999.06). Splitting large data sets into smaller subsets does not provide a significantly better fit to the data. We thus conclude that on a timescale of one month, the images are generally not strongly affected by surface variability. However, \kisr\ from Stokes I profiles show a dependence with the length of the data set. The smallest \kisr\ is obtained at epoch 1998.93 (covering 4 nights only), and the highest one corresponds to epoch 2001.96, (covering as much as 37 nights). Grouping data from epochs 1998.93 and 1999.06 data sets yields a \kisr\ of 1.0 and 2.0 for the magnetic and brightness images respectively, i.e. 11\% and 74\% above the \kisr\ one can reach when separating both epochs (the better stability of the magnetic \kisr\ can be attributed to the weaker constraint provided by polarized profiles).    

Moreover, the two Stokes I profiles plotted in Fig. \ref{fig:2profiles}, taken at rotation phases as close as 0.11\% of a rotation cycle, but observed 20 stellar rotations apart, show clear differences (reaching as much as 9\% of the line depth) between both observations (while all profiles obtained around phase 0.71 are very similar in shape, see Fig. \ref{fig:dec98}). The blue part of the latest profile shows a bump corresponding to a new-born spot located at phase 0.8 and latitude 40\degr\ (Fig. \ref{fig:jan99}), while the red wing of the profile is deeper, owing to the disappearing of the polar spot appendage pointing toward phase 0.55 in the 1998.93 image. These differences demonstrate that surface variability can no longer be ignored on a timescale of order of 2 months.

\begin{table*}
\caption[]{Characteristics of the reconstructed images. The first and second columns list the epoch and timespan of the data sets. The third and fifth columns respectively list the  reduced \kis\ associated to the brightness and magnetic images, while the fourth and sixth columns represent the percentage of spot and the field strength integrated over the stellar surface. }
\centerline{\begin{tabular}{cccccc}
Epoch  & \# of nights&  \kisr   & spot & \kisr & B$_{\rm int}$\\
  &  & Stokes I & (\%)  &   Stokes V &  (G)\\
\hline
1998.14            & 30 & 1.2 & 8.2 & 1.0 & 118 \\
1998.93            & 4  & 1.1 & 4.6 & 0.9 & 38  \\
1999.06            & 19 & 1.2 & 7.4 & 0.9 & 73  \\
1998.93 \& 1999.06 & 58 & 2.0 & 4.1 & 1.0 & 79  \\
2000.14            & 35 & 1.4 & 8.1 & 1.2 & 75  \\ 
2001.96            & 37 & 1.4 & 6.7 & 1.3 & 113 \\ 
\hline
\end{tabular}}
\label{tab:kisr}
\end{table*}

\subsection{Long-lived surface structures}
\label{sect:constant}

\begin{figure*}  
\centerline{\psfig{file=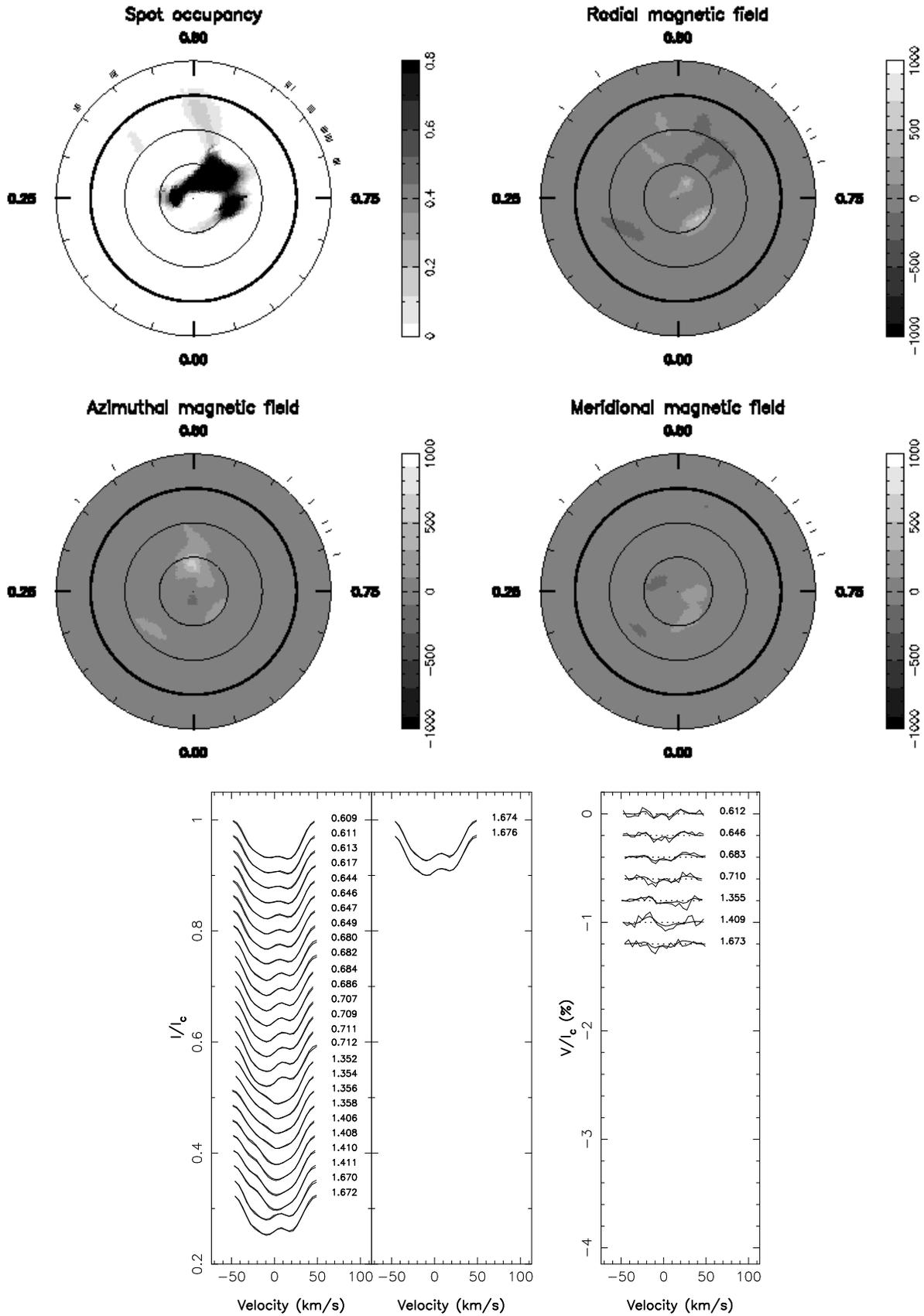,height=13cm}}
\vspace{5mm}  
\centerline{\mbox{\psfig{file=specI_dec98.ps,height=9cm,angle=270}  
\hspace{2mm}      \psfig{file=specV_dec98.ps,height=9cm,angle=270}}}  
\caption[]{Same as Fig \ref{fig:feb98} for 1998.93 data set.}  
\protect\label{fig:dec98}  
\end{figure*}  

\begin{figure}  
\centerline{\psfig{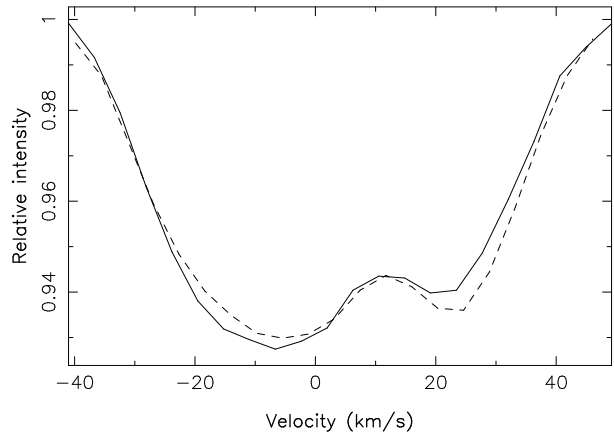}}  
\caption[]{Mean Stokes I profiles obtained at phase 0.7125 on 1998 Dec 05 (full line) and phase 0.7136 on 1999 Jan 31 (dashes). The difference between both observations reaches 9\% of the line depth.}  
\protect\label{fig:2profiles}  
\end{figure}  

\begin{figure*}  
\centerline{\psfig{file=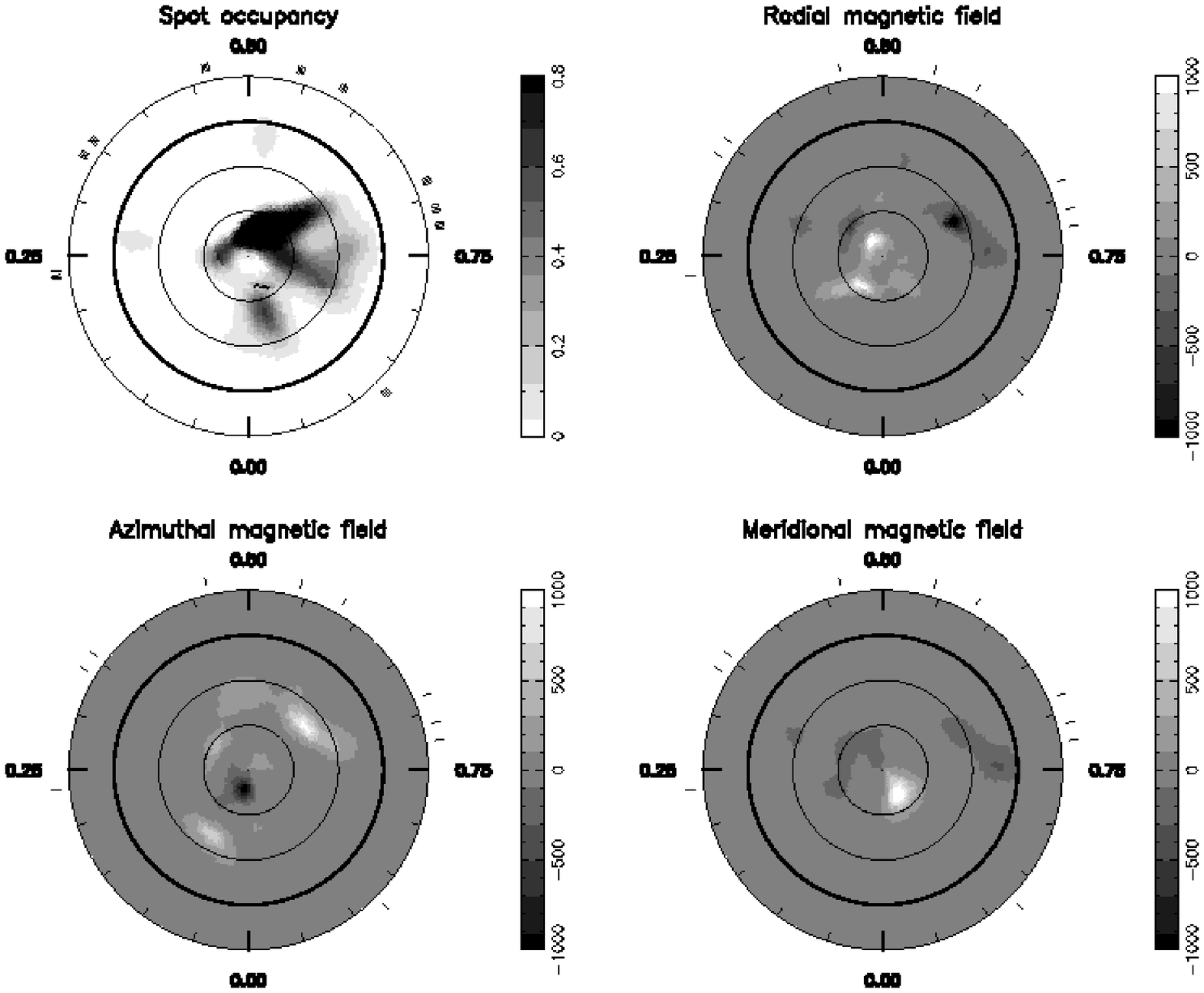,height=13cm}}  
\vspace{5mm}  
\centerline{\mbox{\psfig{file=specI_jan99.ps,height=9cm,angle=270}  
\hspace{2mm}      \psfig{file=specV_jan99.ps,height=9cm,angle=270}}}  
\caption[]{Same as Figure \ref{fig:feb98} for the 1999.06 data set.}  
\protect\label{fig:jan99}  
\end{figure*}  

\begin{figure*}  
\centerline{\psfig{file=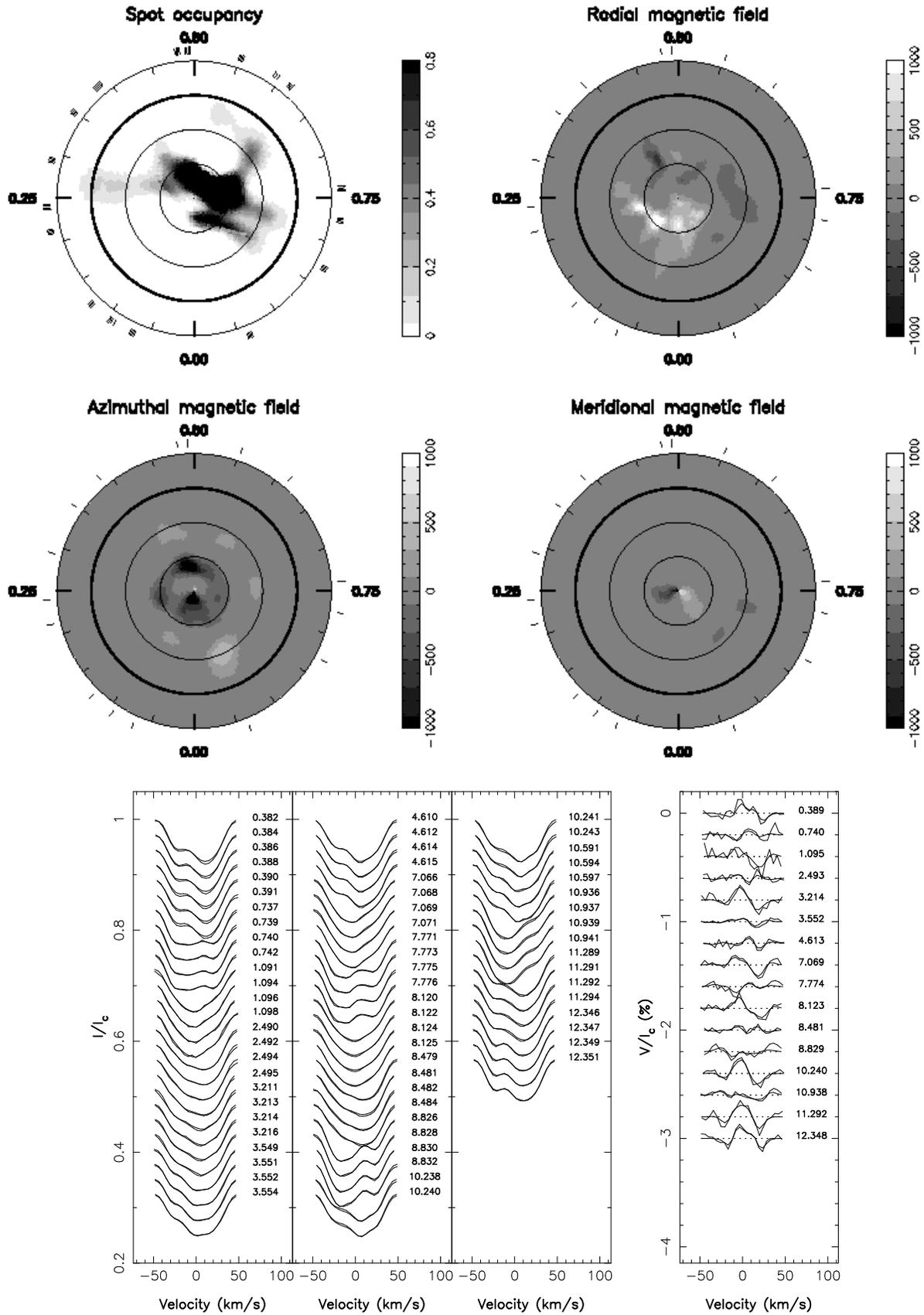,height=13cm}}  
\vspace{5mm}  
\centerline{\mbox{\psfig{file=specI_feb00.ps,height=9cm,angle=270}  
\hspace{2mm}      \psfig{file=specV_feb00.ps,height=9cm,angle=270}}}
\caption[]{Same as Figure \ref{fig:feb98} for the 2000.14 data set.}  
\protect\label{fig:feb00}  
\end{figure*}

\begin{figure*}  
\centerline{\psfig{file=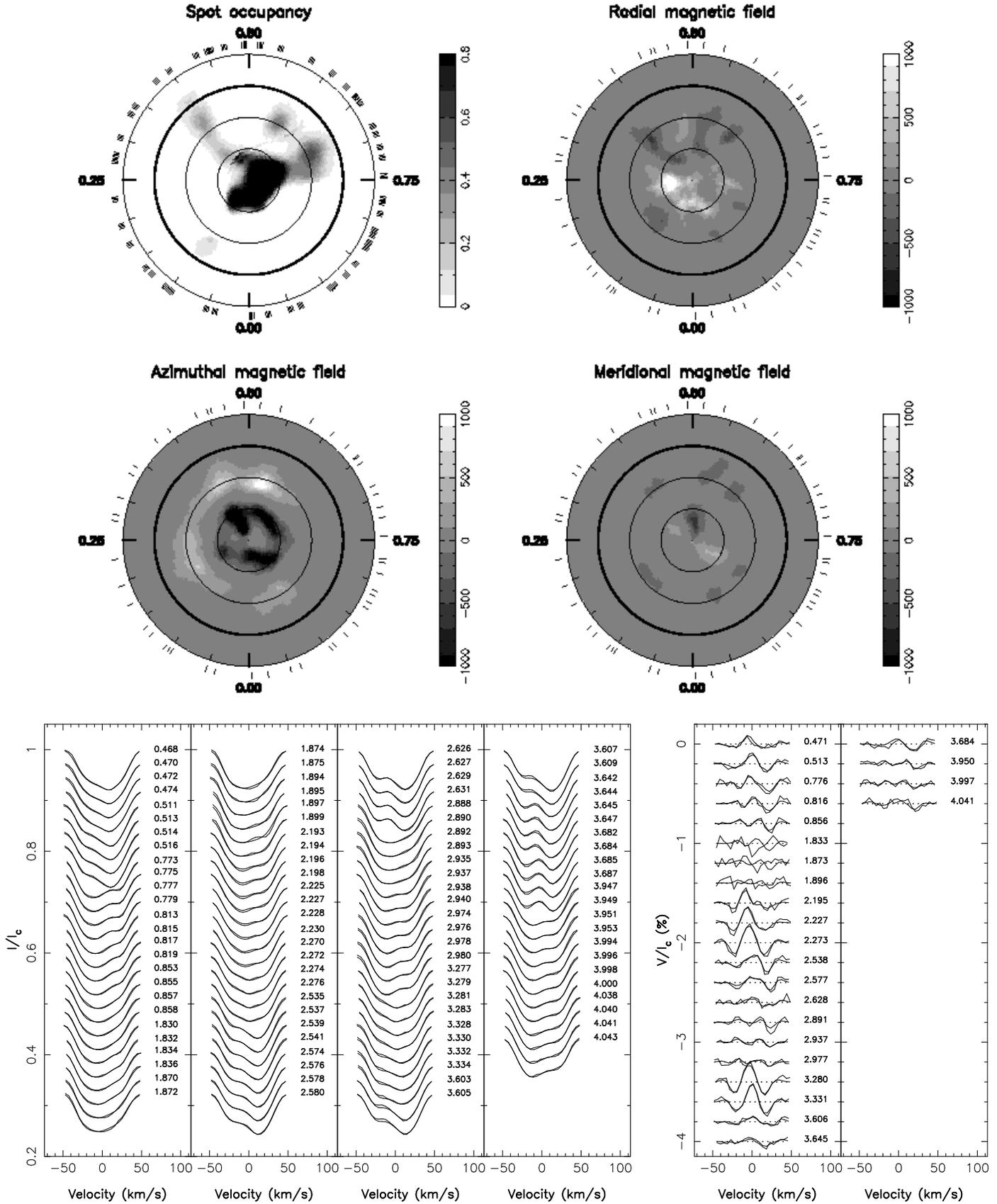,height=13cm}}  
\vspace{5mm}  
\centerline{\mbox{\psfig{file=specI_dec01_epoch1.ps,height=9cm,angle=270}  
\hspace{2mm}      \psfig{file=specV_dec01_epoch1.ps,height=9cm,angle=270}}}
\caption[]{Same as Figure \ref{fig:feb98} for the 2001.96 data set.}  
\protect\label{fig:dec01}  
\end{figure*}  

\begin{figure*}
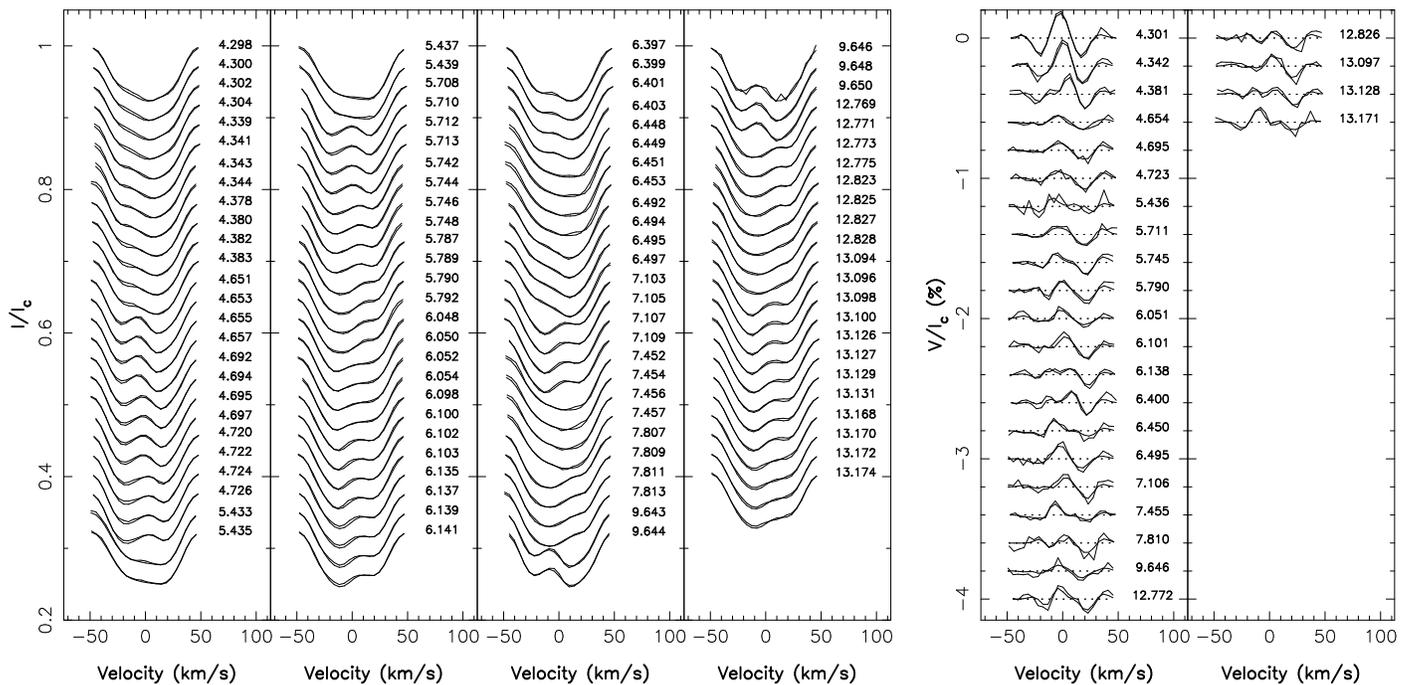
  
\addtocounter{figure}{-1}
\centerline{\mbox{\psfig{file=specI_dec01_epoch2.ps,height=9cm,angle=270}  
\hspace{2mm}      \psfig{file=specV_dec01_epoch2.ps,height=9cm,angle=270}}}
\caption[]{(continued)}  
\protect\label{fig:spec_dec01_1}  
\end{figure*}  

The magnetic images of HR~1099 reported in Fig. \ref{fig:feb98}, \ref{fig:dec98}, \ref{fig:jan99}, \ref{fig:feb00} and \ref{fig:dec01} show most of the characteristics already outlined by Donati et al. (1992), Donati (1999) and D03a. 

The most obvious surface feature in the brightness images is the large, high contrast 
polar spot appearing in all reconstructed images. This feature is never centered on the pole, but is usually concentrated higher than latitude 60\degr. Its shape can be rather 
complex, for instance at epoch 1998.14, when its main component was located at 
phase 0.55 and latitude 60\degr, with several satellite spots forming a partial ring around the pole. Apart from the high-latitude spot, several smaller spots or groups of spots (as in Fig. \ref{fig:dec01}) appear at lower latitude (close to the equator) on all images. Cross-correlating every latitude rings of brightness images obtained at different epochs does not show evidence of any particular longitudinal dependence of the low-latitude structures occupancy. 

Some magnetic features we discuss hereafter have been repeatedly observed over 13 years since the first images of Donati et al. (1992). These structures, detected through several instrumental setups, are also consistently reproduced with different inversion techniques (see the magnetic image of \hr\ obtained with spherical harmonics decomposition from the Stokes V data set of epoch 1998.14 presented here, assuming either an unconstrained or a linear combination of force-free fields, Donati 2001).
  
In particular, we note the persistent 
presence of magnetic regions in which the field is mostly horizontal (i.e. parallel to the surface). These regions show up as two distinct 
azimuthal field regions of opposite polarity. The first of these regions is a ring of counter clockwise field confined at a latitude of about 30\degr. This ring appears fragmented in all images (partly due to the effect of incomplete phase coverage in the case of magnetic topologies associated with epochs 1998.93 and 1999.06). The latitude of the ring does not evolve between 1998.14 and 2001.96. We also note that the ring is never perfectly axisymmetric, but rather appears as a string of spots of intense magnetic field.    

The second region with horizontal field is a ring of clockwise azimuthal field above latitude 60\degr. At two epochs (2000.14 and 2001.96), the 
ring draws a well-defined circle centered on the pole, with a field intensity in excess of 1 kG. On the other images, this magnetic region is reconstructed as a bipolar pair centered on the pole, both in the
azimuthal and meridional components of the field. As explained by Donati (1999), 
such bipolar pairs are only a visual artifact produced by the projection of the field in spherical coordinates, reflecting a single magnetic structure of horizontal and roughly homogeneously oriented field, passing through the pole. On our images, the center of the ring is steadily located between phases 0.5 and 0.6.

The other magnetic regions are mainly radial field spots. Their distribution is 
more complex than the horizontal field regions and show weaker correlation from 
one epoch to the next. Global trends in their distribution over the stellar photosphere can be distinguished though. If we consider the longitudinally averaged radial and azimuthal components of the algebraic magnetic field as a function of latitude (i.e. their contribution to the axisymmetric component of the large-scale field, Fig. \ref{fig:field_dist}), we clearly observe a predominance of positive radial field above latitude 60\degr, most spots of negative polarity being confined at lower latitudes. Moreover, the latitude limit between these two regions roughly corresponds to the transition region between both rings of azimuthal magnetic field. These conclusions are fully consistent with those of D03a. We also confirm that the axisymmetric component of the magnetic topology is very stable over years, as can be deduced from comparison of Fig. \ref{fig:field_dist} with a similar plot in Donati (1999).
  
\subsection{Short term evolution of surface structures}
\label{sect:short}

The time interval between TBL and AAT observations never exceeds 2 months. The comparison of the two data sets gives us the first opportunity for studying in detail the short term evolution the structures at the surface of \hr.

First of all, the observations at epoch 2001.96 (2001.99 at the AAT) represent the first bi-site simultaneous spectropolarimetric observations of HR~1099. Comparing the resulting images constitutes the most accurate consistency check one can perform. The main difference between both data sets is the shorter timespan of AAT observations (covering about one third of the observing time spent at TBL at the same epoch). We first note that all large magnetic and brightness structures, i.e. the rings of horizontal field, the radial field regions, the high-latitude starspots all feature a location, shape and intensity that agree very well in both images. Of particular interest is that the similarity still holds for low-latitude structures as well, demonstrating that reconstruction biases (usually stronger at such latitudes) are essentially insignificant in our case. 

On the other hand, brightness images show at least one clear discrepancy for one spot located at latitude 20\degr and phase 0.40. This spot, appearing as a large and contrasted feature in the TBL image, shows up only as a weak blob in the AAT image. An image (not shown here) reconstructed from a subset of our data selecting only spectra secured within the AAT observing window confirm that these differences are actually due to photospheric variability acting within less than two weeks. Differences are also visible in the details of the polar spot, suggesting that small structures may have a lifetime of order of only a few weeks.

The same kind of evolution is detected at other epochs as well, with similarity between AAT and TBL images decreasing for an increasing time gap between observations. The comparisons are however limited by the difference in phase coverage between TBL and AAT images, which restrict such considerations to high latitude features, or to regions well-monitored with both instruments. 

To the relatively short lifetime of the smallest structures in the brightness and radial field images, we can contrast the longer term stability of larger patterns. In fact, the only striking example of a large-scale reorganization of the photospheric magnetic field within a short time is visible at epoch 2000.14, where the large, intense azimuthal field spot located at phase 0.57 in the 1999.97 AAT image seems to have almost completely vanished from the TBL image, obtained two months later. Another such difference is that the overall field strength inside the low latitude azimuthal field ring is significantly smaller at epoch 2001.96 than at 2001.99.

\subsection{Long term evolution of surface structures}

\begin{figure}  
\centerline{\psfig{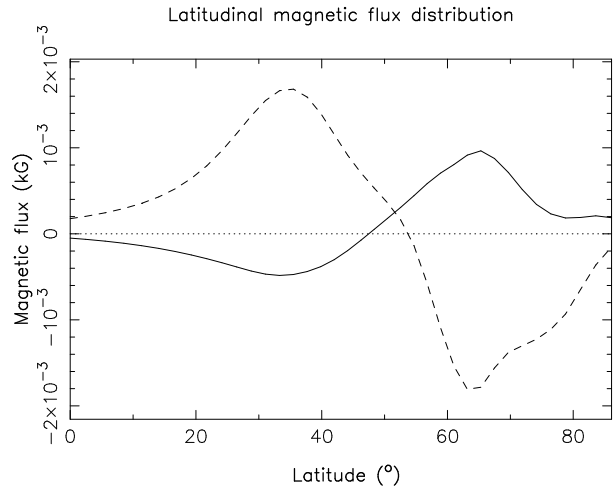}}  
\caption[]{Latitudinal distribution of the radial (full line) and azimuthal ( dashed line) components of the magnetic field, averaged over all observing epochs.}  
\protect\label{fig:field_dist}  
\end{figure}  

As already emphasized, the axisymmetric component of the field topology is remarkably stable over years. No change as obvious as a global polarity switch has yet been observed in the succession of images presented in this study, nor in any other similar ones (Donati 1999, D03a). In particular, the latitudinal location of the rings of azimuthal magnetic field does not significantly evolve from one epoch to the next. However, our observations confirm the trend pointed out by D03a, who reported that the relative fraction of the magnetic energy contained inside the horizontal component of the field was fluctuating with time. From our observations (assuming that the meridional field should be counted as part of the radial component for latitudes less than 45\degr, owing to the crosstalk problem mentioned in Sect. \ref{sect:procedure}), we derive a fraction equal to 79\% and 56\% at epochs 1998.14 and 2000.14 respectively (77\% when averaged over all observing epochs), thus showing that the toroidal component of the field usually contains most of the magnetic energy, except at epoch 2000.14, when both components were of equal importance.

From cross-correlating the latitude rings between brightness images obtained at different epochs (assuming a solid-body rotation of the stellar surface with a rotation period of 2.83774 d), we observe a regular angular shift of the high-latitude spot toward increasing phases. This trend is also readily visible on the brightness images themselves, as well as on those of D03a. It is also compatible with previous images (Donati 1999) except for that corresponding to epoch 1991.96. This regular phase shift of the polar spot is associated to a beat period of about 10 - 12 years. This apparent rotation of persistent polar features can also be tracked from the high-latitude radial field features. However, we must keep in mind that the presence of the very dark brightness polar region is likely to influence the location of the reconstructed magnetic field at high-latitude, therefore introducing a potential spurious correlation between the locations of brightness and radial field structures close to the pole.

\section{Differential rotation}  
\label{sect:diffrot}  

\subsection{Principle}

\begin{table*}
\caption[]{Surface rotation parameters derived from HR~1099 TBL observations. For each epoch, \omeq\ (equatorial rotation rate) and \dom\ (difference of rotation rate between equator and pole) are listed for the reconstructed brightness profiles (columns 2 and 3) and for the associated Stokes V profiles (columns 4 and 5). Line 4 gives the parameters derived from a data set obtained by grouping 1998.93 and 1999.06 profiles sets, and the last line provides the \drot\ law derived when grouping the 2001.96 data set obtained at the TBL with the 2001.99 observations secured by Donati et al. (2002a) at the AAT.}
\begin{tabular}{ccccc}
      & \multicolumn{2}{c}{Brightness images} & \multicolumn{2}{c}{Magnetic images} \\
\multicolumn{1}{c}{Date}  &   \omeq   &   \dom & \omeq   &   \dom \\
year  &   \rpd    &  \mrpd & \rpd    &  \mrpd \\
\hline
1998.14 & $2.22018 \pm 0.0018$ & $10.6 \pm 3.1$ & $2.2464 \pm .0059$ & $40.9 \pm 9.2$ \\
1998.93 & $2.209 \pm 0.033$  & $15.6 \pm 39.8$  & -- & --\\
1999.06 & $2.2241 \pm 0.0045$ & $21.8 \pm 9.2$ & $2.219 \pm 0.011$ & $18.9 \pm 25.6$ \\
1998.93 \& 1999.06 & $2.2172 \pm 0.0029$ & $10.6 \pm 4.4$ & $2.2201 \pm 0.0032$ & $24.1 \pm 6.4$ \\
2000.14 & $2.1915 \pm 0.0096$ & $-3.8 \pm 14.3$ & $2.233 \pm 0.011$  & $28.6 \pm 17.2$ \\
2001.96 & $2.2206 \pm 0.0009$ & $13.2 \pm 1.7$ & $2.2269 \pm 0.0023$  & $17.0 \pm 3.8$ \\ 
2001.96 \& 2001.99 & $2.2241 \pm 0.0004$ & $15.21 \pm 0.82$ & $2.2232 \pm 0.0009$ & $9.68 \pm 1.8$ \\
\hline
\end{tabular}
\label{tab:diffrot}
\end{table*}

\begin{figure}  
\centerline{\psfig{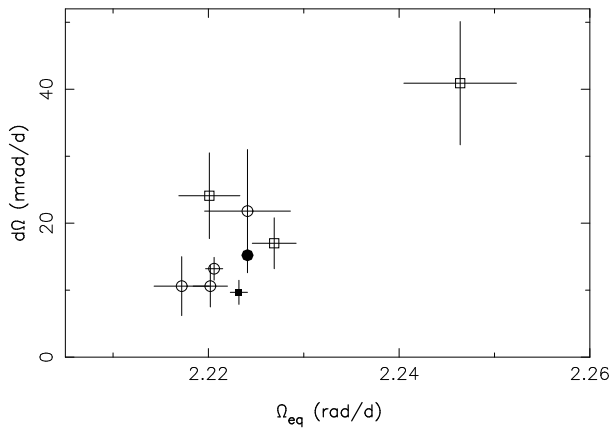}}  
\caption[]{Estimates of \drot\ parameters obtained with TBL data sets. The circles (resp. squares) represent estimates derived from Stokes I (resp. Stokes V) profiles. The plotted data points are only those with error bars smaller than 10 \mrpd. The black-filled symbols represent the parameters obtained when grouping 2001.96 and 2001.99 data sets (i.e. TBL and AAT observations). Note that the error bar corresponding to the Stokes I profiles of epoch 2001.96 \& 2001.99 is smaller than the symbol width.}  
\protect\label{fig:plot_diffrot_TBL}  
\end{figure}

Owing to the rotation period of HR~1099 (very close to the orbital period of 2.83774~d), single-site observations of this star cannot provide data sets with dense phase sampling over one single rotation cycle. At the latitude of Pic du Midi for instance, observations of HR~1099 can hardly cover more than 15\% of a rotation cycle per night, with almost redundant phase observation every 3 days, except for a small shift of 5.7\% of a rotation cycle. A very dense phase sampling of HR~1099 thus requires 18 days in theory, and often more than one month in practice. It implies at the same time that our images of HR~1099 cannot be considered as snapshots and that any kind of temporal variability (like large-scale surface flows) are likely to distort the surface patterns during data collection. In this section, we present an estimate of the photospheric shear at the surface of \hr\ induced by differential rotation.

First of all, we note that the restrictive observing conditions inherent to HR~1099 make inappropriate the techniques commonly used for estimating the surface rotational shear, based on cross-correlating two successive images of the star. As mentioned in Section \ref{sect:short}, images obtained at distant epochs are not correlated, except for the location of the permanent polar spot. No information about \drot\ can thus be extracted from such image comparison. The second possibility would be to split our data sets into smaller subsets and compare the successive images obtained from each subset. However, as pointed out by Petit et al. (2002), the induced reconstruction biases are in this case likely to dominate the differential rotation signal.

We therefore choose the dedicated method of Petit et al. (2002), which proves to be well-adapted to the case of intermediate rotators like HR~1099. This method is based on the reconstruction of brightness and magnetic images for various sets of \drot\ parameters (\omeq\ and \dom, see Equation \ref{eq:diffrot}) in order to find the parameter pairs that minimize the information content of the reconstructed images.

\subsection{Differential rotation measurements}

The \drot\ parameters derived from our sets of brightness and magnetic profiles are listed in Table \ref{tab:diffrot} and shown in Fig. \ref{fig:plot_diffrot_TBL}. We first note that a rotational shear (evaluated by the \dom\ parameter) is firmly detected in our Stokes I data sets (to within as much as $7\sigma$ at epoch 2001.96) as well as in our Stokes V data sets (to within $4\sigma$ at epochs 1998.14 and 2001.96). The parameters derived from some epochs are less accurate, and several data sets did not provide any reliable estimates, with error bars greater than a few tens of \mrpd. Not surprisingly, the most accurate sets of profiles are those collected on the largest timespan, therefore displaying the largest shifts of surface structures during data collection. A high quality level of the data is also necessary, as can be illustrated by the poor result derived from the Stokes I profiles of epoch 2000.14, mostly reflecting the rather low \sn\ of this data set.

However, expanding the timespan above the lifetime of photospheric structures increases the risk of obtaining spurious \drot\ detection, produced by aliases between tracers appearing at close-by locations at the stellar surface. This effect is already present when grouping 1998.93 and 1999.06 Stokes I profiles. The newly born low latitude spot group located around phase 0.75 at epoch 1999.06 (Fig. \ref{fig:jan99}) can indeed be mistaken for the disappearing spot visible at phase 0.5 at epoch 1998.93 (Fig. \ref{fig:dec98}), erroneously suggesting a strongly anti-solar \drot\ (i.e. with a stellar equator rotating slower than the pole). This obviously spurious solution is indeed suggested by our measurements as a secondary \kisr\ minimum, in addition to the \drot\ parameters reported in Table \ref{tab:diffrot} (and equal to $2.1934 \pm 0.0059$ \rpd\ and $-24.6 \pm 8.8$ \mrpd\ for \omeq\ and \dom\ respectively). This effect, that can obviously be tracked down in the data themselves in this particular case, can occur in a much more subtle way in other situations, and generate systematic biases of the reconstructed \drot\ parameters. This example tells us that we cannot derive reliable parameters from data sets collected on a timescale longer than the lifetime of the small surface structures, i.e. typically 4 weeks. 

Most observations secured at the AAT are not close enough to our own observations to guarantee that surface variability will not degrade the surface \drot\ estimates. The only epoch for which AAT and TBL data can be safely grouped is the bi-site campaign of 2001.96 (TBL) and 2001.99 (AAT). The extremely dense data set thus generated (308 Stokes I profiles and 77 Stokes V profiles altogether) constitutes an excellent base to derive a \drot\ law, with both an optimal timespan of 37 nights and a high \sn. The resulting set of parameters is consistent with independent estimates derived from TBL and AAT subsets alone, providing the most accurate estimate of a surface shear ever achieved on a fast rotator, with an uncertainty on \dom\ of the order of 1~\mrpd only.

Finally, we note that the \drot\ parameters we derive do not show evidence for strong time-variability, as reported by, e.g. D03b on younger objects. We speculate that this essentially results from the large relative error bars of our estimates. Only one of the estimates (corresponding to the 1998.14 Stokes V data) significantly differs from the other values (see Fig. \ref{fig:plot_diffrot_TBL}), with a 2.75$\sigma$ discrepancy for \omeq, but with \dom\ consistent with other measurements. Further estimates with smaller error bars are needed to confirm that such a variability can indeed arise from a physical process.
  
\section{Discussion}  
\label{discuss}  

We report evidence that the surface of \hr\ primary star is differentially rotating, with a lap-time (time for the equator to lap the pole by one complete cycle) of the order of 480 days. This result does not confirm previous studies of Vogt et al. (1999) and Strassmeier \& Bartus (2000), reporting that the \drot\ of \hr\ is antisolar, i.e. with its pole rotating faster than its equator. We emphasize however that such studies were carried out with low \sn\ and sparse phase sampling. Moreover, we insist on the fact that the result presented here is repeatedly derived from completely independent data sets obtained at different epochs (from 1998.14 to 2001.96), with different instruments (MuSiCoS and UCLES), and consistently recovered both from brightness and magnetic images. Moreover, the difference of rotation rate between the pole and the equator we derive in the present work ($\dom=15.2 \pm 0.8$ \mrpd) is in agreement with the period variations observed in photometric studies ($\dom=13.1$ \mrpd, Henry et al. 1995).

The most intriguing result of our study is that the rotational shear we measure is significantly weaker than those previously estimated on other active fast rotators (Donati \& Cameron 1997, Donati et al. 2000, Barnes et al. 2000 who measured laptimes 4 to 12 times shorter than in the present study). Moreover, the lap-time of \hr\ primary star is 4 times smaller than that of the Sun. The first reason we may invoke to explain this discrepancy is the evolutionary stage of \hr's primary component (a subgiant). As reported in a companion paper (Petit et al. 2002b), the differential rotation on the G5 FK~Com (sub)giant HD~199178 is again of the same magnitude  as that of the Sun (i.e. about 4 times stronger than that of \hr), despite its similarly deep convective zone. Sadly enough, HD~199178 cannot be taken as a strict analog to \hr, for example because of its higher mass (1.65 M$_\odot$). Such a difference may indeed partly explain its stronger surface shear, as suggested by other observational studies (Donati et al. 2000, Reiners \& Schmitt 2002). However, this dependence of \drot\ on stellar mass was detected on young dwarfs, and may not hold for more evolved objects with very deep convective zones. The effect that may most likely weaken the surface shear of \hr\ is the strong tidal forces that it suffers, imposing very efficiently co-rotation not only to the binary system itself, but also to the rotation within the convective zones of both stars. A possible consequence of these tidal forces is to limit the amount of the surface shear. The theoretical work of Scharlemann (1981, 1982) shows that the differential rotation of an evolved component of a RS CVn system is indeed likely to be strongly weakened (though not totally suppressed), in very good agreement with our observations. In the case where the shear persists, the stellar envelope is still forced to co-rotate in average, which means that part of the envelope rotates faster than the system, the other part slower, and that there is a co-rotating latitude. In our case, this latitude of co-rotation is equal to 50\degr. 

Applegate (1992) predicts that the fluctuations of the orbital period of the system (monitored as the conjunction phase $\phi_0$) may be related to an exchange between kinetic and magnetic energy of the convective zone during the magnetic cycle, and may also show up as temporal variations of the \drot. Owing to the fact that small error bars were only achieved from the bi-site campaign of 2001.96 \& 2001.99, we have not yet detected definite variations of the parameters, and are not yet able to confirm or contradict this prediction. However, the expected \dom\ fluctuations required by this mechanism (from $\dom \approx 0$ to $\dom \approx 40$ \mrpd, D03b) are still roughly compatible with our results and make this issue a promising prospect for future observations. 

As reported by D03b, the angular momentum $J$ within a stellar convective zone may be simply related to the rotation parameters measured at the surface of the star in the case of simple models of the velocity field within the convective envelope~:

\begin{equation}
J(\omeq, \dom) \propto \omeq - \lambda \dom
\label{eq:fast}
\end{equation}

\noindent where $\lambda$ depends on the assumed internal rotation model and on the internal stellar structure. In the particular case where the velocity field is supposed to be close to that of the Sun (angular rotation rate not depending on the distance from the centre of the star), $\lambda$ is equal to 0.2. For the velocity fields expected for very rapid rotators (angular rotation constant over axisymmetric cylinders), $\lambda$ becomes dependent on stellar internal structure, and is equal to 0.68 in the case of \hr\ (D03b). In the case of \hr, one can safely consider that tidal forces are strong enough to impose average co-rotation within the convective zone. This means in particular that \omeq\ will be equal to the orbital rotation rate $\Omega_0 = 2,21415$ \rpd\ whenever \drot\ is suppressed within the convective zone, i.e. that $J(\omeq, \dom) = J(\Omega_0, 0)$. Using the most accurate values of \omeq\ and \dom\ we have estimated (from the Stokes I and V data sets secured at epoch 2001.96 \& 2001.99), the values of $\lambda$ we derive from this relation are equal to $0.7 \pm 0.1$ and $0.9 \pm 0.2$ for Stokes I and V data respectively. This result suggests that the angular velocity field inside the deep convective envelope of \hr\ is close to that expected for rapid rotators.

The weakness of the surface differential rotation of \hr\ is of course a major constraint when investigating the dynamo processes generating its magnetic topology. The most interesting feature of the reconstructed topology is the presence of a strong azimuthal component of the magnetic field at the surface of the star, which may be related to the large-scale toroidal component of the dynamo field. In the case of the Sun, the large-scale toroidal field is believed to be confined at the interface layer between the radiative core and the convective envelope. The fact that it reaches the photospheric level may therefore suggest (as was already pointed out in several articles, see e.g. D03a) that the dynamo may be partly distributed in the convective envelope, or at least significantly active close to the surface. Moreover, the regions of azimuthal field are close to axisymmetry (at least when averaged on several years). This behavior is more obvious on \hr\ than on any object of the sample of young active stars for which magnetic topologies have been reconstructed up to now. As mentioned by D03a, the deep convective zone of \hr's primary allows the maintenance of a strong dynamo activity despite the rather low \drot. In any case, the detection of a well-structured surface azimuthal field (along with a definite differential rotation) suggests that an $\alpha\Omega$ dynamo is operating in this star, implying that the numerical simulations of dynamo processes in tidally-interacting stars carried out by Moss \& Tuominen (1997) (assuming that there is no \drot\ in the stellar convective zone and that only $\alpha^2$-type dynamos can be generated) may not apply here. 

Our study of short-term and long-term variability of \hr\ shows that surface changes are significant for the smallest structures on a timescale of only a few weeks. This typical evolution time holds both for brightness and magnetic spots. Such a variability clearly shows that \drot\ measurements reported by Vogt et al. (1999) (based on comparison of images secured several months apart) are likely to suffer from the aliasing problems that we tried to avoid in the present study. The largest axisymmetric structures (the large polar spot and the rings of azimuthal magnetic field) show a much longer term stability (longer than a decade). The long lifetime of most surface structures on \hr\ (longer than for AB~Dor, Donati \& Cameron 1997) may be related to the smaller surface shear of \hr.

Finally, we note that the present observations of \hr\ primary star magnetic topology do not indicate major changes in the magnetic topology of this star during the last 10 years. In particular we do not detect, as on LQ~Hya (Donati 1999, D03a) any trend announcing a global polarity switch of the magnetic patterns. However, we confirm the discovery of D03a who pointed out that the relative weights of the poloidal and toroidal components of the field are varying with time. The toroidal component of \hr\ field is always dominating the magnetic flux, except at epoch 2000.14, when the large-scale toroidal and poloidal fields contained a same level of energy. The monitoring of such fluctuations may provide some important tests for future stellar dynamo theories.

\section{Conclusions and prospectives}  
\label{conclusion}  

This study, showing evidence for a weak solar-like differential rotation at the surface of the primary K1 subgiant of the RS CVn system \hr\ (with a laptime of order of 480d), gives preliminary indications of the impact of tidal forces on the convective zones of late-type components of close binary systems. 

However, we critically lack \drot\ measurements on a sample of single evolved fast rotators to prove that such a weak surface shear is indeed due to a tidal torque, rather than to other stellar parameters, such as the depth of the convective zone or the stellar mass. We have already started this systematic investigation with the study of the FK~Com G5 (sub)giant HD~199178 (Petit et al. 2002b), and we will soon pursue our investigations with FK~Com itself. We are also carrying out the same type of study on other RS CVn systems (UX~Ari, II~Peg, $\sigma$~Gem) in order to investigate if a weak surface differential rotation is a general feature of close late-type binaries.

Finally, we plan to pursue our monitoring of the above-mentioned active stars with the new generation spectropolarimeters (ESPaDOnS at the CFHT and NARVAL at the TBL), to benefit from increased spectral resolution and \sn, and from the opportunity to conduct the multi-site studies that prove to be necessary to obtain accurate estimates of \drot\ in intermediate rotators. Such high-quality data, along with long-term monitoring engaged with the present article, should make it possible to test Applegate's (1992) prediction that \drot\ of RS~CVn systems may vary during a magnetic cycle, and may be connected with the observed fluctuations of the orbital periods of these objects. 

\section*{ACKNOWLEDGMENTS}

GAW, JDL, SLSS, SS and TAAS acknowledge grant support from the Natural Sciences and Engineering Research Council of Canada (NSERC).

\end{document}